\documentclass[a4paper]{jpconf}
\usepackage{lmodern}        
\usepackage[T1]{fontenc}
\usepackage[utf8]{inputenc} % input encoding
\usepackage{graphicx}  %to include figures
\usepackage{amsmath}
\usepackage{amssymb}
\usepackage{stmaryrd}
\usepackage{mathrsfs}
\usepackage{mathabx}
\usepackage{upgreek}
\usepackage{comment}
\usepackage{verbatim}
\usepackage[pdftex,bookmarksopen,bookmarksnumbered,linktoc=all,hidelinks]{hyperref}

\begin{document}
\title{Study of dynamical systems and large-scale structure}

\author{Dumiso Mithi$^{1}$, Saikat Charkraborty$^{2,1}$, Shambel Sahlu$^{1,3}$, and Amare Abebe$^{1,4}$}

\address{$^1$ Centre for Space Research, North-West University, Potchefstroom 2520, South Africa}
\address{$^2$ The Institute for Fundamental Study “The Tah Poe Academia Institute”,  Naresuan \\
 University, Phitsanulok 65000, Thailand \\}

\address{$^3$ Department of Physics, Wolkite University, Wolkite, Ethiopia}
\address{$^4$ National Institute for Theoretical and Computational Sciences (NITheCS), South Africa}

%only put the email of the main corresponding author, not of all authors.
\ead{\url{mithidumiso@gmail.com}}
%\lesssim$ 1 Mpc
\begin{abstract}
In this study, we employ dynamical systems methods to analyse the large-scale structure by considering two distinct interaction models (linear and non-linear) within the dark sector, associated with a specific dynamical dark energy model inspired by the Veneziano ghost theory in quantum chromodynamics (QCD). In these models, the dark energy density ($\rho_{DE}$) varies with the Hubble parameter ($H$), expressed as $\rho_{DE} = \alpha H + \beta H^2$. After defining the dimensionless parameters, we present autonomous equations that allow us to find the trace $\text{Tr}(J)$ and the determinant $D(J)$. With these solutions, we demonstrate the presence of unstable, saddle, and stable fixed points, corresponding to the radiation-, matter-, and dark-energy-dominated eras, respectively. Our results suggest that these models are theoretically viable for representing the interaction between dark sector fluids.

%Additionally, this work assesses how these models agree with recent cosmic measurements particularly the luminosity distance measurement through supernova survey data, which helps explain the universe's accelerating expansion. The best-flues of the cosmological parameters are calculated through Markov Chain Monte Carlo (MCMC) simulations for these cosmological models, enabling the presentation of Hubble and distance modulus magnitude diagrams for a comprehensive analysis of the accelerating universe.
\end{abstract}

\section{Introduction}
%In standard cosmology, the widely accepted and successful model is the Lambda Cold Dark Matter (\(\Lambda\)CDM) model, which aligns remarkably well with current cosmological observations. Its success comes from its ability to make theoretical predictions that preceded the observations and to explain a wide range of cosmological phenomena across various scales exceeding previous models, such as the Einstein-de Sitter model. Theoretical predictions, such as the temperature structure of cosmic microwave background radiation, and the model can explain large-scale observations, such as the large-scale structure of the universe
In modern cosmology, the Lambda Cold Dark Matter (\(\Lambda\)CDM) model is the most widely accepted framework for understanding the universe's large-scale structure, composition, and evolution. Its significant success and widespread support stem not only from its consistency with contemporary observations but also from its precise predictions of phenomena like the large-scale distribution of galaxies and clusters, driven by the evolution of initial density perturbations through gravitational instability in a universe dominated by dark matter \cite{2003RvMP...75..559P, peebles1993principles,2001MNRAS.328.1039C}, even before these phenomena were observed
\cite{2024PDU....4501524M,2024arXiv240602672L,2022PhLB..82736955N,hobson2006general}. However, the model struggles to fully explain the fundamental properties of its key components: dark matter, dark energy, and inflation. Alongside persistent issues like the $H_0$ tension, fine-tuning, and the coincidence issues, new challenges like the $S_8$ tension have emerged \cite{2021MNRAS.507.3473B,2020PhRvD.101f3531T,2024arXiv240708427T}. 

Standard cosmology faces limitations in both early- and late-time contexts, prompting cosmologists to explore alternatives such as dynamical dark energy models, interactions within the dark sector, and interactions between dark matter and radiation \cite{mithi}. The limited observational data leave the true nature of dark energy poorly understood, allowing for the possibility that it may be dynamic rather than a simple cosmological constant, prompting researchers to explore alternative models that incorporate dynamic aspects to better explain the universe's accelerated expansion. These models strive not only to describe the physical characteristics of dark energy, but also to tackle issues such as fine-tuning and coincidence problems \cite{mithi,2009PhRvD..80h3511B,2019EPJC...79..659B,2020EPJC...80...55P}. This study employs a dual approach: it looks at a dynamic dark-energy model derived from the Veneziano ghost theory in quantum chromodynamics defined as
\begin{equation}\label{2.8}
      \rho_{DE} = \alpha H + \beta H^2,
\end{equation}
where \(\alpha\) and \(\beta\) are constants with units of (energy)\(^3\) and (energy)\(^2\), respectively. The revised sentence describes a theoretical model of dark energy that aims to resolve some common issues associated with cosmological models, particularly the fine-tuning problem and the stability of the universe \cite{2009PhRvD..80h3511B,2019EPJC...79..659B}. Additionally, it considers only the coupling within the dark sector, where gravitational interactions exist between this modified dark energy and dark matter via an interaction term \(Q(t)\). This term was initially introduced to explain the currently small value of the cosmological constant, but it is also significantly helpful in alleviating the coincidence issue.

While a portion of the pressureless matter in the universe evolves independently without any coupling to dark energy, we approximate this by neglecting this nonzero but relatively small part. Consequently, the total energy density of a flat cosmos is composed of dark matter (\(\rho_m\)), dark energy (\(\rho_{DE}\)), and radiation (\(\rho_r\)). The continuity equations for these energy densities are then given by:
\begin{subequations}
     \begin{align}
     &\dot{\rho}_r + 4H\rho_r = 0,\label{2.9a}\\
     &\dot{\rho}_m + 3H\rho_m = Q(t),\label{2.9b}\\
     &\dot{\rho}_{DE} + 3H(1 + \omega_{DE})\rho_{DE} = -Q(t).\label{2.9c}
     \end{align}
\end{subequations}
In the interaction between dark energy and dark matter, the energy exchange rate is denoted by \(Q(t)\). When \(Q(t)<0\), energy is transferred from dark matter to dark energy; conversely, when \(Q(t)>0\), energy flows from dark energy to dark matter. This interaction influences the equation of state (EoS) of dark energy, \(\omega_{DE}\), potentially shifting its nature from quintessence (\(\omega_{DE}>-1\)) to phantom (\(\omega_{DE}<-1\))  \cite{2017IJMPD..2650098G, 2023PDU....4001211M,2016RPPh...79i6901W}.
%To study the dynamics of the universe, we start with the Friedman equations \cite{coley2003dynamical} and based on Planck's 2018 data, it is suggested that our universe is spatially flat according to the joint Lambda cold dark matter ($\Lambda$CDM) model and baryon acoustic oscillation (BAO) measurements \cite{aghanim2020planck}. It is a fundamental principle that energy in the universe is always conserved. Consequently, the Friedman equation for conserved energy in a spatially flat cosmos can be expressed as follows%Dynamical systems techniques are highly effective for studying cosmological topics, enabling the qualitative analysis of system behaviour near fixed points through linearisation and enhancing our understanding of the temporal dynamics. These methods allow complex nonlinear differential equations to be transformed into a set of first-order nonlinear ordinary differential equations:
\section{Dynamical System Analysis} 
The analysis of dynamical systems in \cite{coley2003dynamical} is governed by the Friedmann and Raychaudhuri equations within the framework of the spatially flat $\Lambda$CDM model, supported by baryon acoustic oscillation data \cite{aghanim2020planck}, and defined as follows:
\begin{subequations}
    \begin{align}
    &H^2 = \frac{8\pi G}{3} \rho_{\text{tot}}, \label{a} \\
    &\frac{\ddot{a}}{a} = -\frac{4\pi G}{3} (\rho_{\text{tot}} + 3p_{\text{tot}}). \label{a1}
    \end{align}
\end{subequations}
Here, $\rho_{\text{tot}}$ represents the total energy density, and $p_{\text{tot}} = \omega_{\text{eff}} \rho_{\text{tot}}$ symbolises the effective pressure. The parameter $\omega_{\text{eff}}$ serves as the effective equation of state (EoS), distinguishing contributions from dark energy, dark matter, and radiation, thus enhancing our understanding of their roles in the universe's evolution \cite{kim2006equation}. Considering all components, (i) if $\omega_{\text{eff}} > 0$, radiation is the dominant factor, leading to the deceleration of the universe due to its positive pressure; (ii) if $\omega_{\text{eff}} \approx 0$, the universe behaves as pressureless matter, indicating a decelerating expansion primarily influenced by dark matter, with minimal roles played by dark energy and radiation; and (iii) if $\omega_{\text{eff}} < 0$, it indicates negative pressure, typically associated with dark energy, which accelerates expansion by counteracting gravitational forces \cite{mithi,2001LRR.....4....1C}.

 The Friedmann equation (\ref{a}) describes the universe's expansion rate in relation to the total energy density, while the Raychaudhuri equation (\ref{a1}) defines the acceleration or deceleration of the universe based on the total energy density and pressure. Dynamical systems techniques simplify complex cosmological equations into first-order systems, allowing for a clearer analysis of system behaviour and dynamics near fixed points through a careful selection of new dynamical variables \cite{coley2003dynamical} such as
\begin{subequations}\label{3.9}
     \begin{align}
     &\Omega_{m}\equiv\frac{8\pi G}{3H^2}\rho_{m}\;,\label{3.9b}\\
      &\Omega_Q\equiv\frac{8\pi G}{3H^3}Q\;,\label{3.9c}\\
     &\Omega_{DE}\equiv \frac{8\pi G}{3H^2}\rho_{DE}=\frac{8\pi G}{3H^2}(\alpha H+\beta H^2)=\frac{8\pi G}{3H}\alpha+\frac{8\pi G}{3}\beta=\frac{8\pi G}{3H}\alpha+\xi\;,\label{3.9a}
     \end{align}
\end{subequations}
where $\xi$ is the early dark energy density parameter that plays a significant role in the dark energy density during the universe's early evolution, while $\frac{8\pi G}{3H}\alpha$ is a late dark energy density parameter that is important for the universe's late evolution \cite{2017IJMPD..2650098G}. By substituting Equation (\ref{a}) into Equation (\ref{a1}) and incorporating the continuity equations using new dynamical variables, we obtain the following equations:

 \begin{subequations}\label{3.10}
\begin{align}
&\omega_{DE}=\frac{2\xi^2+\xi(2\Omega_m+3\Omega_{DE}-3\xi-2)+(\Omega_{DE}-\xi)(\Omega_m+\Omega_{DE}-\xi+2)+2\Omega_Q}{3\Omega_{DE}(\Omega_{DE}+\xi-2)},\label{3.10a}\\
&\omega_{eff}=\frac{5\Omega_{DE}+2\Omega_m-2-3\xi+2\Omega_Q}{3(\Omega_{DE}+\xi-2)},\label{3.10b}\\ 
&q=\frac{3\Omega_{DE}+\Omega_m-\xi-2+\Omega_Q}{\Omega_{DE}+\xi-2}.\label{3.10c}
\end{align}
\end{subequations}
From Equations (\ref{3.10}), the dependence on $\Omega_r$ is eliminated because the dimensionless energy density of a flat universe, $\Omega_m + \Omega_r + \Omega_{DE} = 1$, holds true for all cosmic times. These equations summarise the dynamical history of the universe as it evolves, emphasising that this evolution is governed by the dynamical behaviour of $\Omega_m$, $\Omega_{DE}$, and their interactions. However, as pointed out in \cite{2012CQGra..29w5001A}, cosmologists currently lack a microphysical basis for these interactions. Consequently, they often rely on phenomenological models, which are based solely on observed phenomena or empirical data and do not adequately explain the underlying reasons from the perspective of fundamental physics. Therefore, we adopt the ansatz of a generic form of interaction as:
\begin{equation}\label{6}
    Q=3b^2 H\rho_m^{\delta}\rho_{DE}^{\gamma}\rho_{tot}^{\sigma},
\end{equation}
to investigate the influence of the interaction within the dark sector on the universe's evolution, and given the limited understanding of the properties of dark matter and dark energy, the general form of the interaction term satisfies the conditions for this dimensional analysis \cite{2017IJMPD..2650098G} 
\begin{equation}
   \delta+\gamma +\sigma=1. 
\end{equation}
The variables $\delta$, $\gamma$, and $\sigma$ define the nature of the interaction and are considered integer values for simplicity in dimensions. The coupling constant $b$ is dimensionless, and the interaction form includes $b^2$, which guarantees that $Q$ remains positive; thus, the constant positivity of $Q$ ensures energy transfer from dark energy to dark matter in compliance with the second law of thermodynamics. Moreover, this energy transfer from dark energy to dark matter is essential to solve the coincidence problem \cite{2019EPJC...79..659B,2017IJMPD..2650098G}.
% given that $H$ is positive and standard fluids fulfill these two energy conditions for a perfect fluid \cite{2020CQGra..37s3001K}:  the null energy condition ($\rho_i+p_i\geq0$), and the weak energy condition ($\rho_i\geq0$ and $\rho_i+p_i\geq0$).

The dynamical behaviour of $\Omega_m$ and $\Omega_{DE}$ is determined by differentiating these dimensionless variables with respect to time. Substituting the continuity equations and using new dynamical variables, followed by algebraic manipulations, yields a system of autonomous equations governing the dynamics as
\begin{subequations}\label{3.15}
     \begin{align}
 &\Omega_m^\prime=\frac{{\Omega_{m}}(2\Omega_m+5\Omega_{DE}-3\xi-2)+\Omega_Q(2\Omega_m+\Omega_{DE}+\xi-2)}{(\Omega_{\rm DE}+\xi-2)},\label{3.15a}\\
    &\Omega_{\rm DE}^\prime=\frac{[\Omega_{DE}-\xi][\Omega_m+4\Omega_{DE}-4+\Omega_Q]}{(\Omega_{\rm DE}+\xi-2)},\label{3.15b}
     \end{align}
     \end{subequations}
 where $\Omega_i^\prime=\frac{\Omega_i}{Hdt}$ and the preceding equations fully delineate the dynamics of the universe's energy density and their influence by the cosmological model specified in Equation (\ref{2.8}), along with its possible interaction with dark matter. Furthermore, a system of autonomous equations will become singular, provided that the following criteria is met:
\begin{equation}\label{3.16}
    \Omega_{DE}+\xi=2.
\end{equation}
If $\xi = 0$, this would imply $\beta = 0$ when examining Equation (\ref{3.9c}). However, as noted in \cite{2011PhRvD..84l3501C}, $\beta \neq 0$ is crucial for accounting for contributions to the vacuum energy density from QCD generalized ghost dark energy \cite{2019EPJC...79..659B}. According to E. Ebrainhimi et al. \cite{2016IJTP...55.2882E}, the range of $\xi$ for interactions between dark sector fluids is $0 < \xi \leq 0.430$. Thus, $\xi \neq 0$ and falls within $0 < \xi \leq 0.43$ when considering interactions in the dark sector fluids. Furthermore, based on the Friedmann equation for a flat universe, the fractional energy densities should be constrained within $0 \leq \Omega_i \leq 1$. Therefore, $\Omega_{DE} + \xi \neq 2$ at all cosmic times, ensuring that the system of autonomous equations will not diverge, allowing for a meaningful evaluation of the model's behavior in the presence of potential interactions described by the function $\Omega_Q$.

When analysing a system's dynamics, identifying the fixed (or critical) points is crucial as they represent the system's equilibrium states. Understanding their stability reveals the cosmological epochs they may represent. Stability is determined by examining the eigenvalues, which are derived from the determinant $D(J)$ and the Tr (J) trace of the Jacobian matrix as follows: %, rather than directly calculating the eigenvalues. The eigenvalues provide insight into the nature of the critical points as follows:When analysing a system's dynamics, identifying the fixed (or critical) point is crucial as it represents the system's equilibrium state. Understanding the stability of a fixed point reveals what kind of cosmological epoch it may represent. The analysis of fixed points and their stability is key to understanding a system's behaviour. Eigenvalues provide insight into the nature of the critical point, but stability is determined by examining the eigenvalue solutions. Instead of directly calculating eigenvalues, the determinant and trace of the Jacobian matrix are used, with eigenvalues derived from the trace and determinant as follows:
\begin{equation}
    \lambda = \frac{Tr(J) \pm \sqrt{Tr(J)^2 - 4D(J)}}{2} \;. 
\end{equation}
This expression, as discussed in \cite{perko2013differential}, offers useful information on the eigenvalues of the autonomous Equations (\ref{3.15a} and \ref{3.15b}). Below is a summary of the conclusions drawn:
\begin{itemize}
  \item  If $D(J) <0$, the autonomous equations possess two real eigenvalues of opposite signs, resulting in a saddle point at the origin.
  \item  If $ D (J)$, $Tr(J)^2- 4 D(J)>0$  then there are two real eigenvalues with the same sign as Tr (J) and implies that the autonomous equations have a node at the origin, which is stable if $Tr(J)<0$ and unstable if $Tr(J)>0$.
  \item If $D(J)>0$,  $Tr(J)^2\neq4 D(J)$,  and $Tr(J)\neq0$, then there are two eigenvalues of conjugate complexes, $\lambda=a\pm ib$; this means that the autonomous equations are focused at the origin, which is stable if $Tr(J)<0$ and unstable if $Tr(J)>0$.
  \item If $Tr(J)>0$ and $Tr(J)=0$, then the autonomous equations have a centre at the origin; there are two eigenvalues of the pure imaginary conjugate complex, $\lambda=\pm ib$.
\end{itemize}
\subsection{Linear Interaction: Model I \texorpdfstring{ $Q=3b^2 H\rho_m$}{Lg}}
For the case of dark energy and dark matter had to interact in this manner $Q=3b^2H\rho_{m}=3b^2H\frac{3H^2}{8\pi G}\Omega_{m}$, Equation (\ref{3.9c}) would be simplified to $\Omega_Q=3b^2\Omega_{m}$, which leads to the autonomous equations being:
\begin{subequations}\label{3.31}
     \begin{align}
     &\Omega_m^\prime=\frac{{\Omega_{m}}(2\Omega_m+5\Omega_{DE}-3\xi-2)+3b^2\Omega_{m}(2\Omega_m+\Omega_{DE}+\xi-2)}{(\Omega_{\rm DE}+\xi-2)},\label{3.31a}\\
    &\Omega_{\rm DE}^\prime=\frac{[\Omega_{DE}-\xi][\Omega_m+4\Omega_{DE}-4+3b^2\Omega_{m}]}{(\Omega_{\rm DE}+\xi-2)}.\label{3.31b}
     \end{align}
     \end{subequations}
The expressions for the parameters $\omega_{eff}$ and $q$ are as follows:
\begin{subequations}\label{3.32}
\begin{align}
&\omega_{eff}=\frac{5\Omega_{DE}+2\Omega_m-2-3\xi+6b^2\Omega_{m}}{3(\Omega_{DE}+\xi-2)},\label{3.32a}\\ 
&q=\frac{3\Omega_{DE}+\Omega_m-\xi-2+3b^2\Omega_{m}}{\Omega_{DE}+\xi-2}.\label{3.32b}
\end{align}
\end{subequations}
\noindent
 \begin{table}[ht!]
    \centering
    \resizebox{\textwidth}{!}{ % Resizes the table to fit the page width
    \begin{tabular}{|c|c|c|c|c|c|c|c|}\hline
    & $(\Omega_m,\Omega_{DE})$ & $\omega_{eff}$ & $q$ & $Tr(J)$  & $D(J)$  & Stability\\ \hline\hline
    $A$ & $(0,\xi)$ & $\frac{1}{3}$ & 1 & $3+(1+b^2)>0$ & $2+6b^2>0$ & unstable\\\hline\hline
      $B$ & $(1-\xi,\xi)$ & $-b^2$ & $\frac{1}{2}-\frac{3}{2}(b^2)$ & $\frac{1}{2}(1-9b^2)<0$ & $-\frac{3}{2}(1+2b^2-3b^4)<0$ & saddle\\\hline\hline
    $C$ & $(0,1)$ & $-1$ & -1 & $-7+3b^2<0$ & $-12(-1+b^2)>0$ & stable\\\hline
    \end{tabular}
    }
    \caption[Linear interaction, Case: $Q=3b^2 H\rho_m$]{Linear interaction: $Q=3b^2 H\rho_m$. The signature of the fixed points eigenvalue is concluded from the trace and the determinant by assuming $0<b^2$, $\xi\ll1$. This assumption is explained in the text.}
    \label{tab:6}
\end{table}

\begin{figure}[!ht]
\centering
\includegraphics[width=0.35\textwidth]{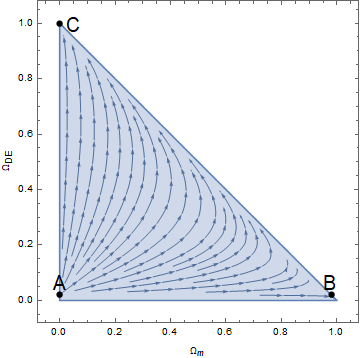}
\includegraphics[width=0.50\textwidth]{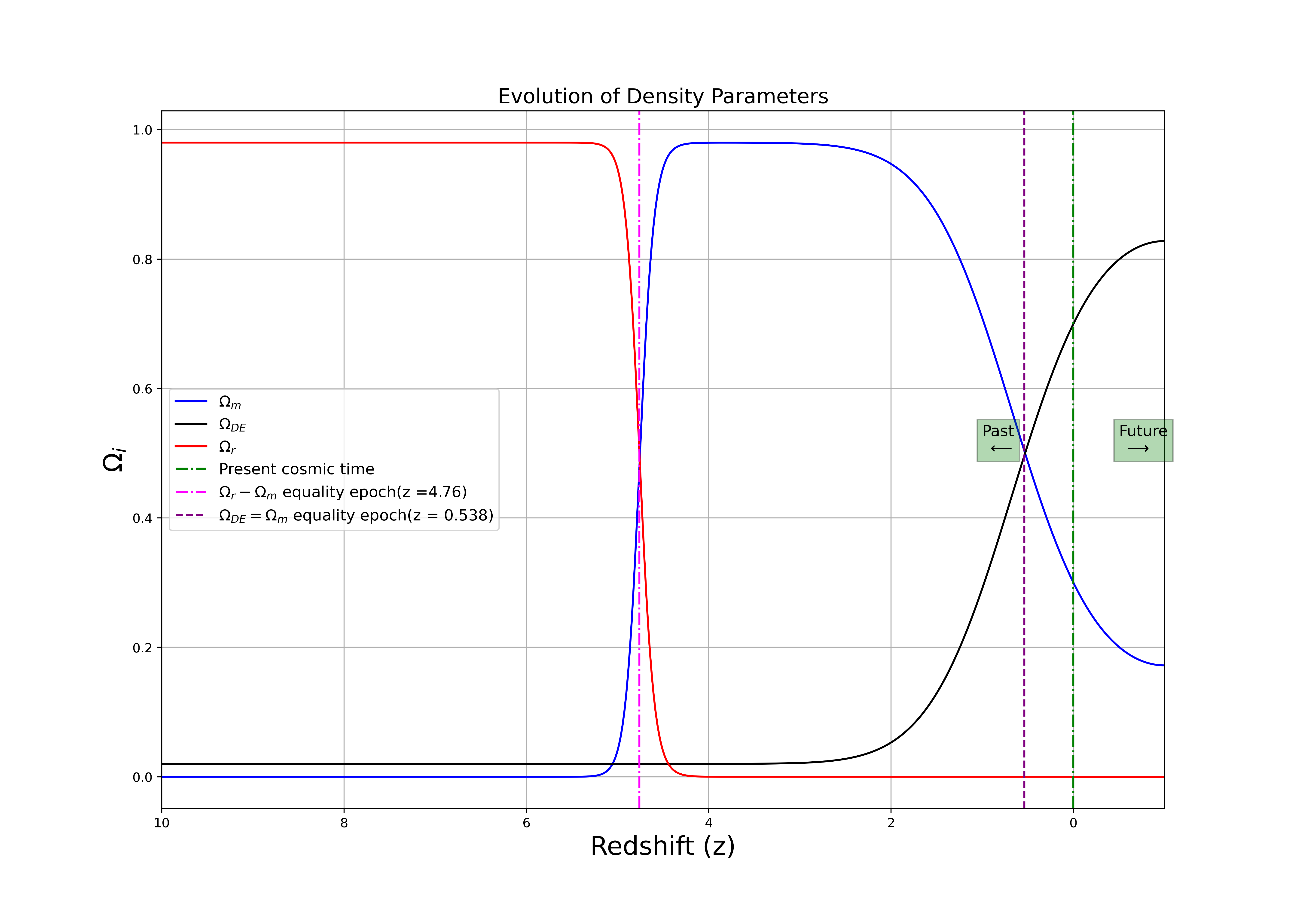}
\caption{\label{fig-sidebyside} (\emph{left}) The phase portrait plot illustrates the trajectories of \(\Omega_{\text{DE}}\) vs \(\Omega_{m}\) for Model I with parameters \((\xi, b) = (0.02, \sqrt{0.40})\). The blue arrows show the direction of the trajectories. Fixed point \(A\) behaves almost as an unstable radiation-dominated era, with all trajectories diverging from it. Fixed point \(B\) is a saddle point behaving as a dark matter-dominated era, where some trajectories approach while others diverge. Fixed point \(C\) signifies a stable dark energy-dominated epoch, where all trajectories converge. (\emph{right}) This plot depicts the evolution of the universe's energy densities as a function of redshift for Model I, using parameters \((\Omega_{m,0}, \Omega_{DE,0}, \xi, b) = (0.315, 0.685, 0.02, \sqrt{0.40})\). The plot tracks the transition from the radiation-dominated era through the dark matter-dominated epoch to the current dark energy-dominated state, all within the viable region consistent with the phase portrait plot.}
\end{figure}

Given \( b^2 > 0 \), when \( \xi < 0 \), the fixed points \( A \) and \( B \) lie outside the viable region, and when \( \xi \approx 1 \), all fixed points correspond to the dark energy era, which is unacceptable. The acceptable range is \( 0 < \xi \ll 1 \), predicting that fixed points \( A \), \( B \), and \( C \) behave almost as radiation-, dark matter-, and dark energy-dominated epochs, respectively. For \( B \) to represent dark matter, the deceleration parameter should be near 0.5, implying \( b \ll 1 \) and \( 0 < b^2 \ll 1 \). Thus, the trace and determinant solutions can be simplified to obtain the expected eigenvalue stability, as shown in Table \(\ref{tab:6}\). Since this model was not constrained by any cosmological data, the results were derived using \( I_{0} = (\Omega_{m,0}, \Omega_{DE,0}, \xi, b) = (0.315, 0.685, 0.02, \sqrt{0.40}) \), where \( (\Omega_{m,0}, \Omega_{DE,0}) \) are from the Planck 2018 Collaboration \cite{2020A&A...641A...6P} and \(\xi, b\) are assumed values since \( 0 < \xi, b \ll 1 \). The phase space and evolution of the density parameters in Figure (\ref{fig-sidebyside}) are based on Equations (\ref{3.31a}) and (\ref{3.31b}).

\subsection{Non-linear Interaction: Model II \texorpdfstring{ $Q=\frac{3b^2H\rho_{DE}\rho_m}{\rho_{tot}}$}{Lg}}
In the case where dark energy and dark matter had to interact in this manner, $Q=3b^2H\frac{\rho_{m}\rho_{DE}}{\rho_{tot}}=3b^2H\frac{\frac{3H^2}{8\pi G}\Omega_{m}\frac{3H^2}{8\pi G}\Omega_{DE}}{\frac{3H^2}{8\pi G}\Omega_{tot}}$, Equation (\ref{3.9c}) would be simplified to $\Omega_Q=3b^2\Omega_{m}\Omega_{DE}$, which leads to the autonomous equations being:
\begin{subequations}\label{3.33}
     \begin{align}
     &\Omega_m^\prime=\frac{{\Omega_{m}}(2\Omega_m+5\Omega_{DE}-3\xi-2)+3b^2\Omega_{m}\Omega_{DE}(2\Omega_m+\Omega_{DE}+\xi-2)}{(\Omega_{\rm DE}+\xi-2)},\label{3.33a}\\
    &\Omega_{\rm DE}^\prime=\frac{[\Omega_{DE}-\xi][\Omega_m+4\Omega_{DE}-4+3b^2\Omega_{m}\Omega_{DE}]}{(\Omega_{\rm DE}+\xi-2)}.\label{3.33b}
     \end{align}
     \end{subequations}
The expressions for the parameters $\omega_{eff}$ and $q$ are obtained as:
\begin{subequations}\label{3.34}
\begin{align}
&\omega_{eff}=\frac{5\Omega_{DE}+2\Omega_m-2-3\xi+6b^2\Omega_{m}\Omega_{DE}}{3(\Omega_{DE}+\xi-2)},\label{3.34a}\\ 
&q=\frac{3\Omega_{DE}+\Omega_m-\xi-2+3b^2\Omega_{m}\Omega_{DE}}{\Omega_{DE}+\xi-2}.\label{3.34b}
\end{align}
\end{subequations}

\begin{figure}[!ht]
\centering
\includegraphics[width=0.35\textwidth]{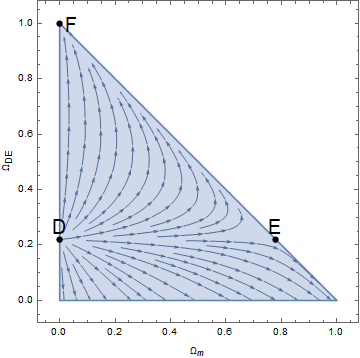}
\includegraphics[width=0.50\textwidth]{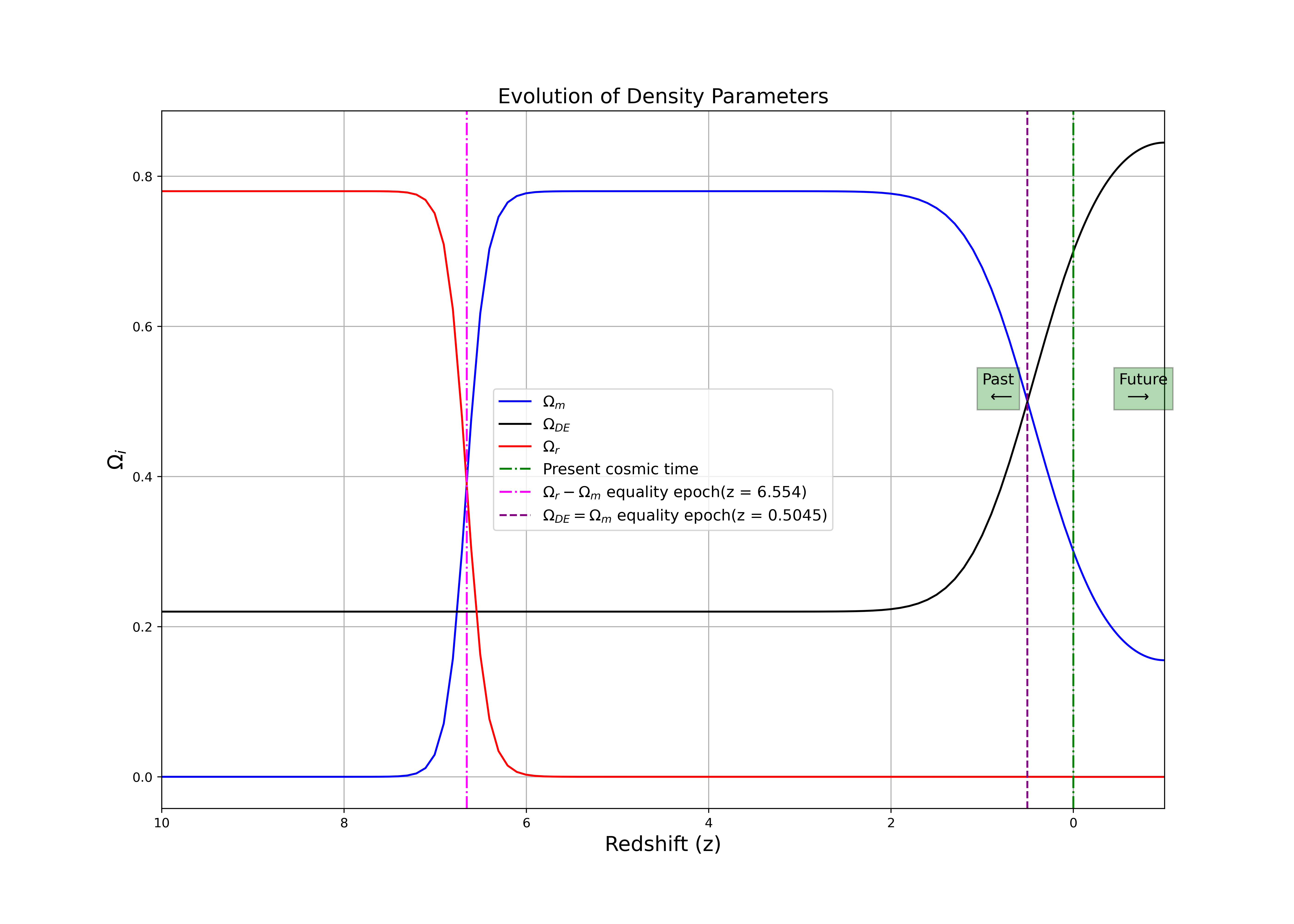}
\caption{\label{fig-sidebyside1} \emph{(left)} The phase portrait for Model II with parameters \((\xi, b) = (0.219, \sqrt{0.27})\) shows the trajectories of \(\Omega_{\text{DE}}\) vs \(\Omega_{m}\).  The fixed point \(D\) is unstable and behave almost as the radiation-dominated era, with trajectories diverging from it. The fixed point \(E\) is a saddle point that behaves almost as a dark-matter-dominated era, where some trajectories approach it while others move away. The fixed point \(F\) is a stable dark energy-dominated era, where all trajectories converge. \emph{(right)} This plot illustrates the evolution of the universe's energy densities with redshift for Model I, using the parameters \((\Omega_{m,0}, \Omega_{DE,0}, \xi, b) = (0.219, 0.779, 0.22, \sqrt{0.27})\). It shows the transition from the radiation-dominated era, through the dark-matter-dominated phase, to the current dark-energy-dominated state, consistent with the phase portrait plot.}
\end{figure}

\begin{table}[ht!]
    \centering
    \resizebox{\textwidth}{!}{%
    \begin{tabular}{|c|c|c|c|c|c|c|c|}\hline
    & $(\Omega_m,\Omega_{DE})$ & $\omega_{eff}$ & $q$ & $Tr(J)$  & $D(J)$  & Stability\\ \hline\hline
    $D$ & $(0,\xi)$ & $\frac{1}{3}$ & 1 & $3>0$ & $2>0$ & unstable \\ \hline\hline
     $E$ & $(1-\xi,\xi)$ & $-b^2\xi$ & $\frac{1}{2}+\frac{3}{2}(-b^2\xi)$ & $\frac{1}{2}>0$ & $-\frac{3}{2}<0$ &saddle\\ \hline\hline
    $F$ & $(0,1)$ & $-1$ & -1 & $-7+3b^2<0$ & $-12(-1+b^2)>0$ & stable \\ \hline
   
    \end{tabular}
    }
    \caption[Non-linear interaction, Case: $Q=\frac{3b^2H\rho_{DE}\rho_m}{\rho_{tot}}$]{Non-linear interaction: $Q=\frac{3b^2H\rho_{DE}\rho_m}{\rho_{tot}}$. The trace and determinant of the eigenvalue of the fixed point $F$ is used to determine its signature provided that $0<b^2$, $\xi\ll1$ is assumed.}\label{tab:3.6}
\end{table}
If $\xi < 0$, the fixed points $D$ and $F$ lies outside the viable region. For $\xi \approx 1$, all fixed points behave as the dark energy era, which is unacceptable. Therefore, the range $0 < \xi \ll 1$ indicates that the fixed points $D$, $E$, and $F$ correspond to radiation, dark matter domination, and  dark energy, respectively. Since $E$ behaves as a dark matter dominated epoch, the deceleration parameter should be $0 < q \leq 0.5$ with $0 < \xi \ll 1$, leading to $0 < b^2 \ll 1$. This simplifies the trace and the determinant for $F$, which yields the expected stability solution, shown in Table (\ref{tab:3.6}). Constrained by E. Ebrainhimi et al. \cite{2016IJTP...55.2882E}, the results were derived using initial conditions fitted to SNIa data: $(\Omega_{m,0}, \Omega_{DE,0}, \xi, b) = (0.219, 0.779, 0.22, \sqrt{0.27})$. The phase space and evolution of the density parameters in Figure (\ref{fig-sidebyside1}) are based on Equations (\ref{3.33a}) and (\ref{3.33b}).

\section{Discussion and Conclusion}
This work extensively analysed the dynamical behaviour of two distinct interacting models. The results indicate that, for both models, the universe transitions from an unstable radiation-dominated state to a saddle matter-dominated epoch, and finally to a stable dark-energy-dominated era, all within a viable region, as illustrated by the phase portraits and energy density plots in Figures (\ref{fig-sidebyside}) and (\ref{fig-sidebyside1}). In their work, Hanif Golchin et al.\cite{2017IJMPD..2650098G} investigated a linear interaction model described by $Q = 3b^2 \rho_{\text{tot}}$,
where \( b^2 \) is the coupling constant, and \( \rho_{\text{tot}} \) denotes the total energy density. They argued that such a linear interaction scenario negates the existence of a radiation-dominated epoch during early times, with recovery only possible in the case of a non-linear interaction. Specifically, under conditions \( 0 < b^2 \) and \( m^2 \ll 1 \) (where \( m^2 \) represents the early dark energy parameter), they concluded that linear models within ghost dark energy theory are cosmologically infeasible.

In contrast, our research reveals that the interaction model $Q = 3b^2 \rho_m,$ where \( \rho_m \) is the dark matter density, offers a different solution. We show that, under the same conditions, \( 0 < b^2 \) and \( \xi \ll 1 \) (where \( \xi \) represents the early dark energy parameter used in our framework), the universe evolves smoothly, preserving appropriate energy densities and allowing for a radiation-dominated epoch. Thus, while linear models may generally face challenges, our model remains viable and provides a robust framework for describing cosmological dynamics within ghost dark energy theory. In conclusion, both Models I and II evolve within a theoretically valid region, consistent with the constraints of cosmological parameters \cite{2016MNRAS.460.1270D}. The theoretical success of these models suggests the potential for a valid interaction within dark sector fluids; however, observational consistency with cosmological data will be addressed in a future paper. %This finding emphasises that not all linear models are flawed and illustrates the necessity for precise selection of interaction terms in ghost dark energy studies. 

\section*{References}
\bibliographystyle{iopart-num}
\bibliography{iopart-num}

\end{document}